\documentclass[pre,aps,preprint]{revtex4}

\usepackage{amsmath}
\usepackage{amssymb}
\usepackage{times}
\usepackage{epsf}
\usepackage{graphicx}
\usepackage{dcolumn}
\usepackage{bm}
\usepackage{CJK}

\begin{document}

\title{Unconventional thermal cloak hiding an object outside the cloak}


\author{Y. Gao$^{1,2}$ and J. P. Huang$^2$}\email{jphuang@fudan.edu.cn}

\affiliation{$^1$Department of Applied Physics, School of Science, Shanghai Second Polytechnic University, Shanghai 201209, China \\
$^2$Department of Physics, State Key Laboratory of Surface Physics, and Key Laboratory of Micro and Nano Photonic Structures (Ministry of Education), Fudan University, Shanghai 200433, China}



\begin{abstract}

All the thermal cloaks reported in the literature can be used to thermally
hide an object inside the cloak. However, a common limitation of this kind
of thermal cloaks is that the cloaked object cannot feel the external heat
flow since it is located inside the cloak; thus we call these cloaks
``conventional thermal cloaks''. Here we manage to overcome this limitation
by exploiting a class of unconventional thermal cloaks that enable the
cloaked object to feel the external heat flow. Our finite element
simulations in two dimensions show the desired cloaking effect.
The underlying mechanism originates from the complementary effect
of thermal metamaterials with negative thermal conductivities.
This work suggests a different method to design thermal devices where heat conduction can be controlled at will.

\end{abstract}

\maketitle

{\bf Introduction.} - The past decades have witnessed the fact that it is more flexible
for people to manipulate light waves, acoustic waves and even
seismic waves than heat conduction. This may be because such kinds
of waves satisfy the wave equations while heat conduction satisfies
the diffusion equation. Therefore, how to manipulate heat conduction
at will is up to now a challenge. If successful, many potential
applications could become true, for example, in solar collectors and
chip cooling.

With an attempt to freely manipulate heat conduction, in 2008
we~\cite{FanAPL08} started from the form invariant of the thermal
conduction equation in distorted coordinates, and proposed a kind of
thermal cloaks for steady state heat flow (namely, the temperature,
$T$, is not a function of time) by using anisotropic and
inhomogeneous (graded) materials. The thermal cloak can hide an
object inside the cloak from the detection by measuring the
distribution of external heat flux. This behavior is because the
cloak and  object do not affect the distribution of temperature
outside the cloak as if they do not exist. Narayana and
Sato~\cite{NarayanaPRL12} experimentally realized this cloaking
effect by using forty alternating layers of two materials, one of
which has a high thermal conductivity and the other of which has a
low thermal conductivity. In addition, while some theoretical
researchers proposed to design such thermal cloaks by using a
simplified method based on homogeneous materials~\cite{HanSR13},
other theoretical~\cite{GuenneauOE12} or
experimental~\cite{SchittnyPRL13,NarayanaAPL13} researchers also
extended the thermal cloaks from steady state heat flow to unsteady
state heat flow (i.e., $T$ is a function of time). However, all the
reported thermal
cloaks~\cite{FanAPL08,LiJAP10,NarayanaPRL12,HanSR13,GuenneauOE12,SchittnyPRL13,NarayanaAPL13,Ma13,DedeAPL13,YangJPD13,HeAPL13,YuFOP11}
have a common limitation: the cloaked object cannot feel the
external heat flow because it is located inside the cloak. In other
words, the object hidden inside the cloak has to be ``blind''. For
clarity, we call these thermal
cloaks~\cite{FanAPL08,LiJAP10,NarayanaPRL12,HanSR13,GuenneauOE12,SchittnyPRL13,NarayanaAPL13,Ma13,DedeAPL13,YangJPD13,HeAPL13,YuFOP11}
``conventional thermal cloaks''. A similar limitation also appears,
but is solved~\cite{LaiPRL09} in optical/electromagnetic cloaks.
Clearly, if the limitation is overcome in thermal cloaks, the
cloaking effect would be more practicable, especially, in the eye of
the cloaked object. This statement is because the object can then
feel external heat flow but cannot be detected by measuring the
distribution of temperature outside the  region of the cloak
and object. That is, the object itself is no longer ``blind''. For
comparison, we call such thermal cloaks ``unconventional thermal
cloaks''. In what follows, we shall propose a recipe for an
unconventional thermal cloak, which can hide an object outside the
cloak.

{\bf Transformation thermophysics.} - Let us start by investigating the thermal conduction equation. Without loss of generality, we omit
the source term and obtain the steady-state conduction equation for the steady state heat flow,
\begin{equation}
{\bf \nabla}\cdot [-\kappa {\bf \nabla} T]=0~, \label{elec_1}
\end{equation}
where $\kappa$ is a thermal conductivity.
Because conduction equations are invariant in their form under the transformation from an
original (regular) coordinate to a transformed (distorted) one~\cite{MiltonNJP06}, we achieve the thermal
conductivity tensor, $\overleftrightarrow{\kappa^{\prime}}$, of the material in the transformed coordinate~\cite{FanAPL08,LiJAP10},
\begin{equation}
\overleftrightarrow{\kappa^{\prime}}=\frac{\mathbf{M} \kappa_{0}
\mathbf{M}^{t}}{ \rm{det}(\mathbf{M})}~, \label{Jacob}
\end{equation}
in terms of the thermal conductivity in the original coordinate,
$\kappa_{0}$. Here $\mathbf{M}$ is the Jacobian transformation
matrix between the original and distorted coordinates,
$\mathbf{M}^{t}$ denotes the transposed matrix of $\mathbf{M}$, and
$\rm{det}(\mathbf{M})$ is the determinant of $\mathbf{M}$. In
general, eq.~(\ref{Jacob}) enables us to obtain the thermal
conductivity tensor of the material in the coordinates with various
types of transformations.

Now we are allowed to design the unconventional thermal cloak of
interest. In this work, for simplicity, we consider
cylindrical cases with coordinate parameters ($r$, $\theta$, $z$) in
the original coordinate and ($r^{\prime}$, $\theta^{\prime}$,
$z^{\prime}$) in the transformed coordinate; the cloak is
schematically shown in Fig.~1. For our purpose, we attempt to fold a
big annulus into a small one while keeping their common surface
unchanged. In detail, as shown in Fig.~1, the large circle with
radius $r= r_c$ is linearly changed into a small one with $r = r_a$.
Meanwhile, the complementary material is obtained by folding the
large annulus (with radius $r$ satisfying $r_b < r < r_c$) into a
small one ($r_a < r <r_b$). As a result, the
object, which is to be hidden, should be put in Region C (as
indicated in Fig.~1), while  Region B of Fig.~1 corresponds
to the desired thermal cloak. Then it becomes necessary to determine
the material parameters for the cloak occupying Region B (Fig.~1).
For this purpose, besides $\theta^{\prime}= \theta$ and
$z^{\prime}=z$, we consider the following coordinate transformation,
\begin{eqnarray}
&&r^{\prime }=\frac{r_a}{r_c}r~~~ ~(0<r<r_c),\label{0rc}\\
&&r^{\prime}=-\frac{r_b-r_a}{r_c-r_b}(r-r_b)+r_b~~~ ~(r_b<r<r_c).
\end{eqnarray}
Regarding this transformation, the Jacobian
transformation matrix, $\mathbf{M}$,
is given by
\begin{eqnarray}
\mathbf{M} =\left(
\begin{array}{ccc}
\frac{\partial r^{\prime}}{\partial r} & \frac{~~\partial r^{\prime}}{r \partial \theta} & 0\\
\frac{r^{\prime} \partial \theta^{\prime}}{\partial r} & \frac{r^{\prime}\partial \theta^{\prime}}{r\partial \theta} & 0\\
0 & 0 & 1
\end{array}
\right)~.
\end{eqnarray}
So, we obtain the thermal conductivity tensor,
$\overleftrightarrow{\kappa}_{1}$, for Region A (as displayed in Fig.~1),
\begin{eqnarray}
\overleftrightarrow{\kappa}_{1}=\frac{\mathbf{M} \kappa_{0}
\mathbf{M}^{t}}{\mathrm{det}(\mathbf{M})}= \left(
\begin{array}{ccc}
1 & 0 & 0 \\
0 & 1 & 0 \\
0 & 0 & (\frac{r_c}{r_a})^{2}
\end{array}
\right)\kappa_{0}~.\label{eq6}
\end{eqnarray}
Similarly, we obtain the thermal conductivity tensor,
$\overleftrightarrow{\kappa}_{2}$, for Region B (in Fig.~1),
\begin{eqnarray}
\overleftrightarrow{\kappa}_{2}=\frac{\mathbf{M} \kappa_{m}
\mathbf{M}^{t}}{\mathrm{det}(\mathbf{M})}= \left(
\begin{array}{ccc}
\kappa_{2,rr} & 0 & 0 \\
0 & \kappa_{2,\theta\theta} & 0 \\
0 & 0 & \kappa_{2,zz}
\end{array}
\right)~,\label{cm}
\end{eqnarray}
where $\kappa_{m}$ is the thermal conductivity of the material
located in Region C (as shown in Fig.~1), and the three diagonal components are respectively given by
\begin{eqnarray}
\kappa_{2,rr} &=& \frac{r_b (r_c-r_a)\kappa_{m}}{r (r_b-r_c)}+\kappa_{m}, \nonumber\\
\kappa_{2,\theta\theta} &=& \frac{r (r_b-r_c)\kappa_{m}}{-r_a r_b+r_b (r_c+r)-r_c r}, \nonumber\\
\kappa_{2,zz} &=& \frac{(r_b-r_c) [-r_a r_b+r_b (r_c+r)-r_c r]\kappa_{m}}{r (r_a-r_b)^2}. \nonumber
\end{eqnarray}

{\bf Simulation results.} - We are now in a position to perform finite element simulations in
two dimensions by using the commercial software, COMSOL
Multiphysics. Fig.~1 is a schematic graph showing four regions (A,
B, C, and D), which are divided by three radii: $r_a=0.5$\,m,
$r_b=0.7$\,m, and $r_c=0.9$\,m (parameters for simulations). Region
B is located between  $r_a$ and $r_b$, and it denotes the
unconventional thermal cloak of our interest. The object, which is
to be cloaked, is located in Region C between $r_b$ and $r_c$, a
region outside Region B. According to eq.~(\ref{eq6}), both Region A
(with radius $r<r_a$) and Region D (with $r>r_c$) are set to have
the same isotropic and homogeneous material (background material)
with thermal conductivity $\kappa_0=40$\,W/(m$\cdot$K). The temperature is
respectively set to be 400\,K and 300\,K at the left and right
boundaries of the squared simulation area with the side length of
2.5\,m; heat insulation is used for both up and down boundaries.
Fig.~2 shows the simulation results of temperature distribution for
three cases. Details are as follows. Fig.~2(a) displays the
temperature distribution for the pure background material; the
distribution is regular. Then, we replace Region C inside the
background material of Fig.~2(a) with an object of thermal
conductivity $\kappa_m = 400$\,W/(m$\cdot$K); see Fig.~2(b). Accordingly,
when compared to Fig.~2(a), Fig.~2(b) shows that the distribution of
temperature within Region D is significantly affected by the object
(which occupies Region C). Next, we use the cloaking material to
replace Region B of Fig.~2(b) according to eq.~(\ref{cm}), and plot
Fig.~2(c). As a result, Fig.~2(c) shows that the temperature
distribution within Region D ($r>r_c$) turns out to be the same as
that in Fig.~2(a). In other words, the thermal cloak (occupying
Region B) can be used to hide the object (located in Region C) from
the detection by measuring the distribution of external heat flux
(within Region D). This behavior also holds for objects (which
occupy Region C) with arbitrary shapes like squares or triangles
(figures are not shown herein). Clearly the underlying mechanism
lies in the complementary effect of materials within Region B, which
have a graded, anisotropic, negative thermal conductivity (see
Fig.~3). Such materials with negative thermal conductivities are
called ``thermal metamaterials'' since they cannot be found in
naturally occurring materials or chemical compounds. Regarding this
kind of thermal metamaterials, we have to address more. Physically,
a negative thermal conductivity corresponds to the fact that heat is
transferred from regions of low temperature to regions of high
temperature. This process means that, to comply with the second law
of thermodynamics, an external work (say, based on Peltier effects~\cite{KovalevEPL12,BakkerPRL10}) should be performed accordingly.
Indeed, researchers have reported a negative thermal conductivity when
investigating chains of rotors with mechanical
forcing~\cite{IacobucciPRE11}. Nevertheless, if the external work is stable, the negative thermal conductivity obtained should also be stable. In this case, the steady-state conduction equation [eq.~(1)] could be used as what we have done in this work. On the other hand, if the external work is unstable, the corresponding  negative thermal conductivity will also be unstable, which is dependent on time. As a result, the distribution of temperature will evolve with time, and thus an unsteady-state conduction equation should be adopted instead.

{\bf Conclusions.} -
In summary, we have proposed a class of unconventional thermal cloaks.
Such a cloak can hide an object, which is located outside the cloaking shell,
from the detection by measuring the distribution of external heat flux.
A feature of this kind of thermal cloaks is that the cloaked object
itself can feel external heat flow. The feature makes it distinctly
different from the conventional thermal cloaks where cloaked objects
cannot feel external heat flow. This feature mainly originates from
the complementary effect of thermal metamaterials with negative thermal
conductivities. Achieving this feature is at the cost of a specific design of configurations. Our simulations can be extended from two dimensions to
three dimensions as long as material parameters are set appropriately.
Thus, by choosing the complementary effect of thermal metamaterials appropriately,
this work suggests a different way to design thermal devices
where heat conduction can be controlled at will.

{\bf Acknowledgments.} - 
Y.G. is grateful to Dr. W. H. Wang for his help. We acknowledge the
financial support by the National Natural Science Foundation of
China under Grant Nos. 11075035 and 11222544, by the Fok Ying Tung
Education Foundation under Grant No. 131008, by the Program for New
Century Excellent Talents in University (NCET-12-0121), by the
Shanghai Rising-Star Program (No. 12QA1400200), and by the CNKBRSF
under Grant No. 2011CB922004. Y.G. acknowledges the financial
support by Young Teacher Training Program of SMEC under Grant No.
egd11005, by Innovation Program of SMEC under Grant No. 12YZ177, and
by NNSFC under Grant No. 11304195.

\clearpage
\begin{figure*}[t]
\begin{center}
\centerline{\includegraphics[width=0.6\linewidth]{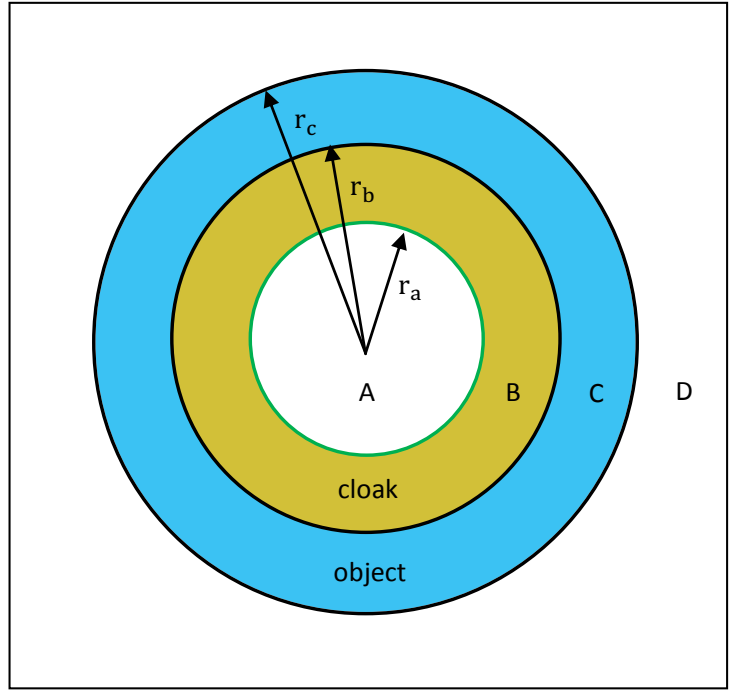}}
\caption{(Color online) Schematic graph showing four regions (A, B,
C, and D), which are divided by three radii: $r_a$, $r_b$, and
$r_c$. The unconventional thermal cloak occupies Region B between
$r_a$ (inner radius) and $r_b$ (outer radius); the object to be
cloaked (or hidden) lies in Region C between $r_b$ (inner radius)
and $r_c$ (outer radius). Others can be found in the main text.
 }
\end{center}
\end{figure*}

\clearpage
\begin{figure*}[t]
\begin{center}
\centerline{\includegraphics[width=0.4\linewidth]{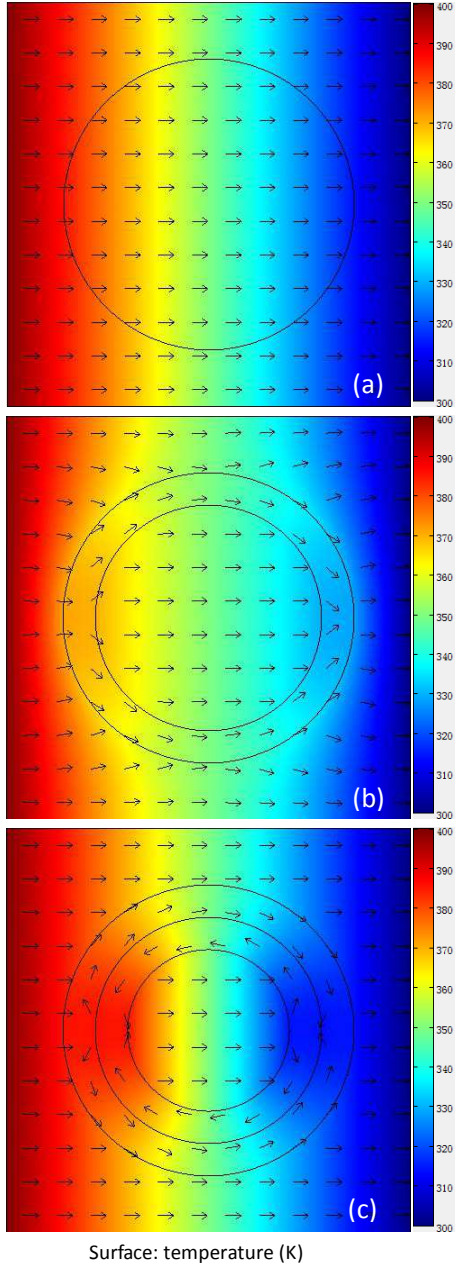}}
\caption{ (Color online) Two-dimensional finite element simulations
of temperature distribution. The color surfaces represent
temperature; the arrows denote the pathway of heat flux. The
temperature is respectively set to be 400\,K and 300\,K at the left
and right boundaries of the squared simulation area with the side
length of 2.5\,m. (a) Temperature distribution of the background
that contains an isotropic and homogeneous material with thermal
conductivity $\kappa_0=40$\,W/(m$\cdot$K). For comparing with (b) and (c),
we add a circle of radius $r = r_c$ to label the position only. (b)
Temperature distribution for the case where an object of thermal
conductivity $\kappa_m=400$\,W/(m$\cdot$K) is used to replace Region C
(between $r_b$ and $r_c$ as indicated in Fig.~1) of the background
displayed in (a). (c) Same as (b), but the cloaking area (Region B)
between $r_a$ and $r_b$ is filled with materials according to
eq.~(\ref{cm}). }
\end{center}
\end{figure*}

\clearpage
\begin{figure*}[t]
\begin{center}
\centerline{\includegraphics[width=0.8\linewidth]{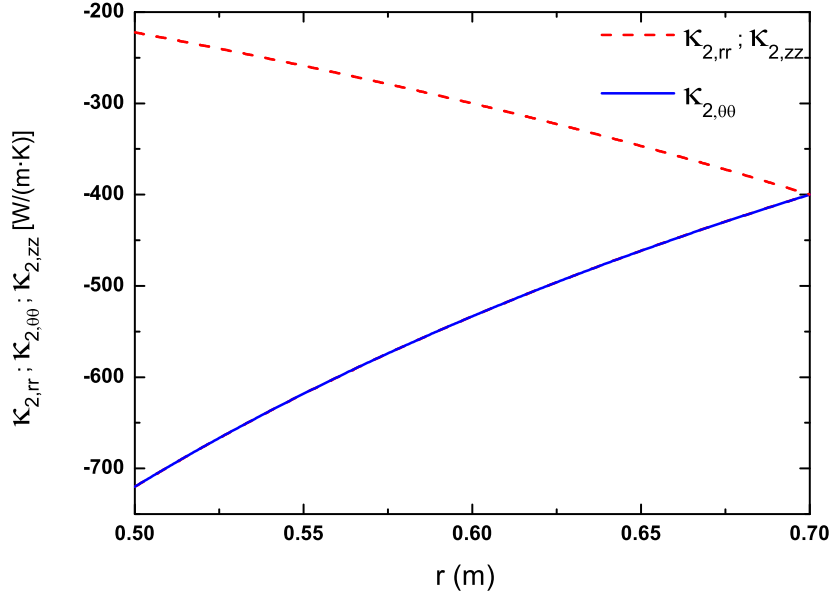}}
\caption{(Color online) The three diagonal components,
  $\kappa_{2,rr}$,
$\kappa_{2,\theta\theta}$ and $\kappa_{2,zz}$, as a function of
radius $r$ within Region B ($r_a < r < r_b$). The upper (red) line
shows the overlapped values of $\kappa_{2,rr}$ and $\kappa_{2,zz}$.
Parameters: $\kappa_m = 400\,$W/(m$\cdot$K), $r_a=0.5$\,m, $r_b=0.7$\,m, and  $r_c=0.9$\,m.
}
\end{center}
\end{figure*}

\end{document}